\documentstyle[12pt,fleqn,epsfig]{article}
\begin{document}
\begin{center}
{\bf INSTITUT~F\"{U}R~KERNPHYSIK,~UNIVERSIT\"{A}T~FRANKFURT}\\
D - 60486 Frankfurt, August--Euler--Strasse 6, Germany
\end{center}

\hfill IKF--HENPG/6--96

\vspace{1cm}

\begin{center}

{\Large \bf Pion Suppression in Nuclear Collisions}

\vspace{1cm}

Marek Ga\'zdzicki\footnote{E--mail: marek@ikf.physik.uni--frankfurt.de}\\
Institut f\"ur Kernphysik, Universit\"at Frankfurt \\
August--Euler--Strasse 6, D - 60486 Frankfurt, Germany\\[0.8cm]

Mark I. Gorenstein\footnote{E--mail: goren@ap3.gluk.apc.org}\\
Bogolyubov Institute for Theoretical Physics\\ 
UK - 252143 Kiev, Ukraine\\[0.8cm]

Stanis\l aw Mr\' owczy\' nski\footnote{E-mail: mrow@fuw.edu.pl} \\
So\l tan Institute for Nuclear Studies,\\
ul. Ho\. za 69, PL - 00-681 Warsaw, Poland \\
and Institute of Physics, Pedagogical University,\\
ul. Le\' sna 16, PL - 25-509 Kielce, Poland
\vspace{1.5cm}

\begin{minipage}{14cm}
\baselineskip=12pt
\parindent=0.5cm
{\small The pion multiplicity per participating nucleon in 
central nucleus--nucleus collisions at the energies 2--15 A$\cdot$GeV
is significantly smaller than in nucleon--nucleon interactions at the 
same collision energy. This effect of pion {\it suppression} is 
argued to appear due to the evolution of the system produced at the
early stage of heavy--ion collisions towards a local thermodynamic 
equilibrium and further isentropic expansion.}

\end{minipage}

\end{center}

\vspace{0.5cm}

\begin{center}
{\it to be published in The European Physical Journal {\bf C} }
\end{center}

\vfill
\today
\newpage

Pions, which are copiously produced in high energy interactions, play 
a key role in the collision dynamics. However, in spite of hard efforts, 
the pion production in nucleus--nucleus interactions is far from being 
completely understood (see, e.g., \cite{St:86}). Thus, it is of 
particular interest to compare the properly normalized data on pion 
multiplicity from nucleus--nucleus and nucleon--nucleon collisions. 

One would expect that inelastic secondary  interactions produce additional 
pions and therefore their number per participating nucleon should be larger 
in nucleus--nucleus than in nucleon--nucleon collisions at the same initial 
energy per nucleon. The experimental data on pion multiplicities amazingly 
contradicts this intuitive expectation. Indeed, it has been recently found 
\cite{Ro:95,Ga:95} for BNL AGS energies and below that the number of produced 
pions per participating nucleon, $\langle \pi \rangle/\langle N_P \rangle$, 
in central collisions of identical nuclei ($A+A$) is lower (pion 
{\it suppression}) than in inelastic nucleon--nucleon ($N+N$) interactions. 

In Fig.~1 we show the ratio $\langle \pi \rangle/\langle N_P 
\rangle$ 
as a function of $\langle N_P \rangle$ at three initial momenta 2.1, 4.5, 
and 15 A$\cdot$GeV/$c$. The data is taken from the compilation \cite{Ro:95},
where the pion multiplicities from various experiments are recalculated
to obtain the total multiplicities independent of the rapidity
and/or transverse momentum cuts. 
In all three cases the relative pion production decreases when going from 
$N+N$ interactions ($\langle N_P \rangle$ = 2) to central $A+A$ collisions. 
At 2.1 A$\cdot$GeV/$c$ the pion yield per nucleon is smaller by a factor of 
about 3. 
At all energies the pion suppression is approximately independent of 
the size of (sufficiently large) colliding nuclei 
(see Fig.~1 and 
the review \cite{Ro:95}). Further, the pion suppression factor defined as  
\begin{equation}\label{supres}
\Delta \frac {\langle \pi \rangle} {\langle N_P \rangle} =
\frac {\langle \pi \rangle_{AA}} {\langle N_P \rangle_{AA}}
- \frac {\langle \pi \rangle_{NN}} {\langle N_P \rangle_{NN}} \;,
\end{equation}
appears to be approximately independent of the collision energy 
(up to BNL AGS energies) \cite{Ro:95,Ga:95}. As seen in Fig.~2, it 
equals about $-0.35$.

The aim of this paper is to discuss the mechanism leading to  pion 
suppression. We try to connect the scaling properties of the suppression 
factor (\ref{supres}) -- its approximate independence of the size of 
colliding nuclei and the initial energy -- with the hypothesis supported
by the existing experimental data that the system created in nucleus--nucleus 
collisions approaches the local thermodynamical equilibrium 
\cite{Sta:95,Ch:96,Cl:97,Yen:97}. 

We assume that the system produced at the early stage of $A+A$ collision 
is formed due to the superposition of $N+N$ interactions. At this stage
the chemical composition of hadronic matter is expected to be the same 
as in the nucleon--nucleon collisions. The system however evolves 
towards  thermodynamic equilibrium and we assume that local 
equilibrium is reached before the system disintegrates into the final state
free hadrons. Then, the difference between the properly normalized 
pion multiplicities in $A+A$ and $N+N$ collisions appears as a result of
\begin{itemize}
\item the chemical
equilibration of the initially nonequilibrium 
hadronic matter, 
\item the hydrodynamic expansion preceding the system freeze--out.
\end{itemize}
We analyse the two mechanisms of pion suppression separately. Let us start 
with the equilibration one. 

A large fraction (50--100 \%) of baryons emitted in inelastic $N+N$ 
interactions is known to be in the state of nucleon isobars (deltas 
and heavier baryonic resonances), which successively decay into pions and 
nucleons \cite{Sh:82,Pi:62,Ei:65,Fi:62,Al:67,Al:68}. On the other hand, 
it is also known, see e.g. \cite{De:87}, that in the equilibrated system 
at temperatures smaller than 150 MeV, which are characteristic for the energy 
domain of interest, the fraction of baryon number carried by deltas does 
not exceed 50 \%. Therefore, when the system created in $A+A$ collisions 
evolves towards equilibrium, the initial surplus of deltas has to be 
reduced causing the suppression of the final state pions. The microscopic 
process responsible for the $\Delta$  absorption is 
$\Delta + N \rightarrow N + N$ \cite{We:90}. Multi--nucleon reactions 
are also discussed in this context, see e.g. \cite{Rz:96}. 
So far the qualitative argument and now we move to a model formulation.

An equilibrium state of the system of nucleons, deltas and pions (heavier 
mesons and baryons are neglected in our considerations) is controlled by two 
parameters: baryon density ($\rho_B$) or baryon chemical potential ($\mu_B$) 
and temperature ($T$). The system, which is formed at the early stage of 
central $A+A$ collision, is assumed to be close to  thermal equilibrium. 
This assumption can be justified by short (relatively to the evolution time)
thermal equilibration time \cite{Go:89,So:96}. The fraction of baryon charge 
carried by deltas, however, exceeds its chemical equilibrium value. Thus, 
we describe such a system in terms of thermodynamics but an additional 
parameter, which measures the delta surplus ({\it DS}), is introduced. 
Specifically, we define
$$
\lambda_{\Delta} \equiv 
{\widetilde\rho_{\Delta} - \rho_{\Delta} \over \rho_B} \;,
$$ 
where $\widetilde\rho_{\Delta}$ and $\rho_{\Delta}$ are the initial 
and equilibrium (corresponding to the initial temperature) densites of deltas. 

Keeping in mind that in the final state there are direct pions and those 
originating from the delta decays, the pion multiplicity per participating 
nucleon is
\begin{equation}\label{mult}
{\langle \pi \rangle \over \langle N_P \rangle} =
{\rho_{\Delta} + \rho_{\pi} \over \rho_B} \;,
\end{equation}
with $\rho_{\pi}$ being the pion density. The suppression factor 
(\ref{supres}) due to the chemical equilibration of initial {\it DS} matter
produced in $A+A$ collisions by a superposition of  $N+N$ interactions then 
reads
\begin{equation}\label{supres1}
\bigg( \Delta  \frac {\langle \pi \rangle} {\langle N_P \rangle}\bigg)_{DS} =
 {\rho_{\Delta}(\mu_B, T) + \rho_{\pi}(T) \over \rho_B}
- {\rho_{\Delta}(\mu_{\Delta}^i,T^i) + \rho_{\pi}(T^i) \over \rho_B^i}\;,
\end{equation}
where $\mu_B$ and $T$ describe the equilibrium state while $\mu_{\Delta}^i$, 
$T^i$ and $\rho_B^i$ the initial nonequilibrium one formed in the early
stage of heavy--ion collision. The particle and energy densities used later 
are given by the well known formulas:
$$
\rho_j(\mu_j, T) =  \int {d^3 p \over (2\pi)^3} \,
{g_j \over e^{\beta (\sqrt{{\bf p}^2 + m_j^2} - \mu_j)} \pm 1} 
\;,\;\;\;\;
\varepsilon_j(\mu_j, T) =  \int {d^3 p \over (2\pi)^3} \,
{g_j \sqrt{{\bf p}^2 + m_j^2} \over 
e^{\beta (\sqrt{{\bf p}^2 + m_j^2} - \mu_j)} \pm 1} \;,
$$
where $m_j$ and $\mu_j$ are particle masses and chemical potentials with
$j = \pi,\; N,\; \Delta$; $\beta \equiv 1/T$. The numbers of internal 
degrees of freedom are: $g_{\pi} = 3$, $g_N = 4$, and $g_{\Delta} = 16$. 
The chemical potential of pions $\mu_{\pi}$ is  zero. 
Due to the strong interaction the above ideal gas formulae are probably 
not very realistic at $T$ larger than, say, 100 MeV and $\rho_B$ 
significantly exceeding normal nuclear density. The most important effect
i.e. the short range repulsion in the hadron system can be taken into
account via the van der Waals correction. Then, the particle 
number ratios, which are of our particular interest,   
could be not far away from their ideal gas values (see e.g. Ref.~\cite{Yen:97}).

The pion density depends 
solely on the temperature while the delta density is a function of the 
temperature and delta chemical potential. The values of chemical potentials 
in Eq.~(\ref{supres1}) are chosen in such a way that
\begin{eqnarray*}
\rho_B   &=&  \rho_{\Delta}(\mu_B,T) + \rho_N(\mu_B,T) \;, \\
\rho_B^i &=&  \rho_{\Delta}(\mu_{\Delta}^i,T^i) + \rho_N(\mu_N^i,T^i) \;, 
\end{eqnarray*}
where $\rho_N$ is the nucleon density. Since we use the parameter 
$\lambda_{\Delta}$ to control the delta surplus, we require that
\begin{equation}\label{lambda}
{\rho_{\Delta}(\mu_{\Delta}^i,T^i) - 
\rho_{\Delta}(\mu_B^i,T^i)\over \rho_B^i} = \lambda_{\Delta} \;, 
\end{equation}
where $\rho_{\Delta}(\mu_B^i,T^i)$ is the equilibrium value of the delta 
density at the temperature $T^i$ and baryon density $\rho_B^i$. 
This equilibrium density is found from the equation
\begin{equation}\label{equi-den}
\rho_B^i =  \rho_{\Delta}(\mu_B^i,T^i) + \rho_N(\mu_B^i,T^i) \;.
\end{equation}

 A complete treatment of the pion suppression requires 
a simultaneous study of collective 
expansion and chemical equilibration processes.
To estimate the role of the two phenomena we however discuss them
separately. Therefore, we assume that the hydrodynamic 
expansion does not develop essentially at the time of chemical
equilibration and the latter process is studied at
the constant volume. Therefore,   
the baryon and energy densities are the same for 
initial and  equilibrium phases:
\begin{eqnarray*}
\rho_B^i=\rho_B \;, 
\end{eqnarray*}
\begin{eqnarray*}
\varepsilon_{\Delta}(\mu_{\Delta}^i,T^i) + 
\varepsilon_N(\mu_N^i,T^i) + \varepsilon_{\pi}(T^i) =
\varepsilon_{\Delta}(\mu_B,T) + \varepsilon_N(\mu_B,T) 
+ \varepsilon_{\pi}(T) \;.
\end{eqnarray*}

After these two equations are solved simultaneously with the 
additional conditions (\ref{lambda}) and (\ref{equi-den}), the suppression 
factor (\ref{supres1}) is a function of three free parameters
choosen  to be
$\rho_B$, $T$ and $\lambda_{\Delta}$. We have calculated numerically 
the pion suppression for the temperatures and baryon densities which cover 
the values characteristic for the hadronic matter formed in $A+A$ collisions. 
The temperature $T$ then varies from 50 to 150 MeV while the baryon density 
$\rho_B$ from $2 \rho_0$ to $5 \rho_0$ with $\rho_0$ = 0.16 fm$^{-3}$ being 
the normal nuclear density.

The pion suppression (\ref{supres1}) disappears when the parameter 
$\lambda_{\Delta}$ goes to zero. To estimate the maximal value of 
$\lambda_{\Delta}$ we note that at high energies of colliding nuclei, 
where the temperature approaches 150 MeV, the equilibrium value of 
$\rho_{\Delta}/\rho_B$ is about 0.5, and consequently  one has 
$\lambda_{\Delta} < 0.5$. At the lowest energies of interest, where 
$\langle \pi \rangle _{NN} /\langle N_P \rangle _{NN} \cong 0.5$, the 
majority of pions in $N+N$ collisions comes from the delta decays while the 
equilibrium delta density in $A+A$ is close to zero. Therefore, we again 
have $\lambda_{\Delta} < 0.5$ and consider $\lambda_{\Delta}=0.5$ as 
the largest value. 

In Fig.~3 
we show the pion suppression as a function of the equilibrium 
temperature $T$ at $\rho_B = 2 \rho_0$ and  $5\rho_0$ for the extreme 
$\lambda_{\Delta}$ value. We also present the results for 
$\lambda_{\Delta}=0.3$. Fig.~3 covers the whole physically reasonable 
domain of $T$, $\rho_B$ as well as $\lambda_{\Delta}$. The suppression 
factor is seen to range from $-0.2$ to $-0.4$. Keeping in mind how strongly
the pion multiplicity varies with $T$, $\rho_B$ and $\lambda_{\Delta}$,
one finds very striking a very weak dependence of the pion suppression on 
the mentioned parameters. This in turn agrees with an approximate 
independence of the suppression factor of the participant number and 
collision energy (cf. Figs.~1,~2). 
More than that, the numerical values 
of the pion suppression due to the equilibration process are close to the 
experimentally measured mean suppression equal of $-0.35$. 

The chemical equilibration leads to an increase of the system
temperature ($T>T^i$), and therefore the number of direct pions
increases as well. The initial nonequilibrium number of deltas 
however strongly decreases. In Fig.~4 we show
$\rho _{\pi}/\rho _B$ and
$\rho _{\Delta}/\rho _B$ 
ratios before (dotted lines) and after (solid lines) the chemical 
equilibration. The sum of these ratios defines the total pion multiplicity
per participating nucleon (\ref{mult}).

The suppression remains almost unchanged when direct pions are removed 
from our calculations by setting $g_{\pi} = 0$. Therefore, the pion 
suppression occurs in our model due to the equilibration of the baryon 
subsystem. Consequently, the assumption that the direct pions are in 
equilibrium is not very important and can be relaxed. We return to this
point at the end of our paper.

 Note also  that the entropy per baryon, 
$\langle S\rangle/\langle N_P\rangle\equiv s/\rho_B$, increases due to the 
chemical equlibration. This is shown in Fig.~5. The entropy density $s$
is calculated from ideal gas formulae for chemical nonequilibrium
initial state with parameters $T^i, \mu_N^i, \mu_{\Delta}^i$
(dotted line) and for chemical equilibrium with parameters
$T, \mu_B$ (solid line).

\vspace{0.5cm}

Let us now estimate the effect of the second mechanism of the pion 
suppression i.e. the absorption of pions due to the system hydrodynamic 
expansion. We call it a delayed freeze--out ({\it DF}) effect in $A+A$ 
collisions: the hadronic system produced in these collisions is larger than 
that from $N+N$ interactions and therefore the freeze--out density is expected 
to be smaller. We 
consider an isentropic evolution of 
the locally equilibrated  hadron matter. It was observed a long time ago 
\cite{Me:78,Gl:88} that number of pions indeed decreases in the course of 
such an expansion of the pion--baryon gas.

The locally equilibrium system starts with $\rho_B$, $T$ and then
expands until the  freeze--out values $\rho_B^f$, $T^f$ are reached. 
The pion suppression then reads
\begin{equation}\label{supres2}
\bigg( \Delta \frac{\langle \pi \rangle}{\langle N_{P} \rangle}\bigg)_{DF}  =\;
{\rho_{\Delta}(\mu_B^f,T^f) + \rho_{\pi}(T^f) \over \rho_B^f}\;
- \; {\rho_{\Delta}(\mu_B,T) + \rho_{\pi}(T) \over \rho_B} \;.
\end{equation}
Since the entropy is assumed to be conserved during the expansion, 
the ratio of the entropy density to the baryon density is constant. 
The freeze--out temperature $T^f$ can be thus expressed as a function
of $\rho_B$, $T$ and $\rho_B^f$. Consequently, the pion suppression 
(\ref{supres2}) is controlled by the same three parameters.

We have found a remarkable `scaling' property of the pion suppression due 
to the isentropic expansion which, as far as we know, has not been noticed 
before. At fixed $T$ the suppression (\ref{supres2}) becomes a function 
of the ratio $\rho_B^f/\rho_B$ only. If the thermal pion contribution to 
the system entropy is neglected, the temperature $T^f$, which  is a
solution of an equation of the isentropic expansion, depends at fixed 
initial temperature $T$ on the ratio $\rho_B^f/\rho_B$ only. 
The delta part of the pion suppression (\ref{supres2}), 
$\rho_{\Delta}(\mu_B^f,T^f)/\rho_B^f\; - \; \rho_{\Delta}(\mu_B,T)/\rho_B$, 
manifests then the same scaling property as $T^f$. The thermal pion break 
the exact scaling, but still the scaling for $T^f$ holds with a high 
accuracy. The thermal pion part of the  suppression (\ref{supres2})
($ \rho_{\pi}(T^f)/ \rho_B^f\; - \; \rho_{\pi}(T)/ \rho_B$) does not
scale, but the effect is numerically small. Therefore, the corrections to 
the `scaling law'  for the total pion suppression factor (\ref{supres2}) 
are very minor (less than a few percent) as long as the initial baryon 
density is sufficiently large, say $\rho_B > 0.5 \rho_0$, and the initial 
temperature is not too big, $T <150$ MeV. 
 In Fig.~6 we show  
$\rho_{\pi}/\rho_B^f$
and
$\rho_{\Delta}/\rho_B^f$
ratios for $T=150$ MeV and $\rho _B=2\rho_0$.
In physical terms, the scaling 
tells us that the pion suppression due to the isentropic expansion mainly 
results from the delta absorption.

Our numerical calculations of the {\it DF} pion suppression (\ref{supres2}) 
are shown as a function of $\rho_B^f/\rho_B$ in Fig.~7. 
One sees that the pion 
suppression due to the expansion is very small for the initial temperature 
$T=50$ MeV and reaches a value of about $-0.4$ at $T=150$ MeV and 
$\rho_B^f/\rho_B = 0.1$. Thus, it is expected to increase with growing 
collision energy (large $T$) and the size of the colliding nuclei 
(small $\rho_B^f/\rho_B$ due to the delayed freeze--out). 

The final pion suppression in $A+A$ collisions combines the effects caused 
by the equilibration and expansion and can be calculated as a simple
sum of the factors (\ref{supres1}) and (\ref{supres2}). As follows from 
the results presented in Figs.~3 and~7, 
the sum varies between $-0.2$ and 
$-0.7$ in the whole physically acceptable domain of the hadronic matter 
parameters. As mentioned above, we expect the pion absorption to increase
with growing size of the colliding nuclei and collision energy. 
This is indeed consistent with data: the independence of the pion suppression
of the size of colliding nuclei breaks down for nuclei as heavy as gold.
Then, the pion suppression (\ref{supres}) equals about $-0.6$  
at 11.6 A$\cdot$GeV$/c$ \cite{Ro:95,Ga:95}. The same trend is found in the
recent GSI SIS results at lower energies \cite{Pe:96}.

The suppression caused by $DS$ and $DF$ mechanisms was calculated
under assumption that direct pion component is in equilibrium i.e. its 
chemical equilibration time is much smaller than the system evolution
time. However, as pointed above, the suppression remains unaffected
when direct pion component is removed from the calculations. This implies 
that our results are valid also for the case when chemical equilibration 
time of direct pions is much larger than the evolution time i.e. the number 
of direct pions is effectively frozeen.
 
We summarize our considerations as follows. The suppression of the pion 
production per participating nucleon, which is observed in central 
$A+A$ collisions at the energies of BNL AGS and below, has been discussed 
within a  thermodynamical approach. 
An approximate independence of the suppression
factor (\ref{supres}) on the collision energy and the participant number
(see Figs.~1 and~2) as well as its numerical value 
agree with 
a scenario of the heavy--ion collision which distinguishes the following
three stages:
\begin{enumerate}

\item The initial preequilibrium stage when the nonequilibrium
hadronic system is formed by a superposition of $N+N$ interactions.

\item The equilibration stage when the number of deltas decreases
to the equilibrium value leading to the reduction of the 
 total number of pions
(see Fig.~3). 

\item The expansion stage of locally equilibrated hot hadronic matter which 
causes an additional pion suppression (see Fig.~7).

\end{enumerate}

For the initial energies of BNL AGS and below
this scenario is checked 
to be qualitatively correct after adding more
mesonic and baryonic resonances and going beyond 
the ideal gas approximation applied here. 
It would be also interesting to check our picture of the pion 
suppression against microscopic transport calculations.  
   
As a final remark we should stress that at the CERN SPS energies 
(160--200 A$\cdot$GeV) one observes a pion {\it enhancement} instead of 
the suppression when going from $N+N$ to $A+A$ collisions \cite{Ro:95}. This 
qualitatively different behaviour can not be 
understood within the 
model presented here.
A novel feature of A+A collisions
at SPS energy is a role of meson
resonances which becomes much more important than
that at AGS energy and below:
a number of mesons at the freeze--out in  
A+A collisions at SPS is several
times larger than number of baryons. 
The chemical equilibration and hydrodynamical expansion in such a system
may lead to a change of the suppression pattern observed at low 
energies
and deserve a special study. 
It is also possible that    
the explanation of the pion 
{\it enhancement} effect requires the introduction of  new mechanisms. 
A formation of the Quark--Gluon Plasma at CERN SPS energies has been 
considered as an obvious candiatate \cite{Ga:95}, however other mechanisms 
are also discussed \cite{Mi:96}.

\vspace{1cm}

{\bf Acknowledgements}. We are indebted to Herbert Str\"obele for discussions
and critical comments on the manuscript. One of us (M.I.G.) expresses 
his gratitude for the warm hospitality at the Institute for Theoretical 
Physics of Frankfurt University when this work was completed. The support 
by BMFT, DFG and GSI is also thankfully acknowledged.


\begin{figure}[p]
\epsfig{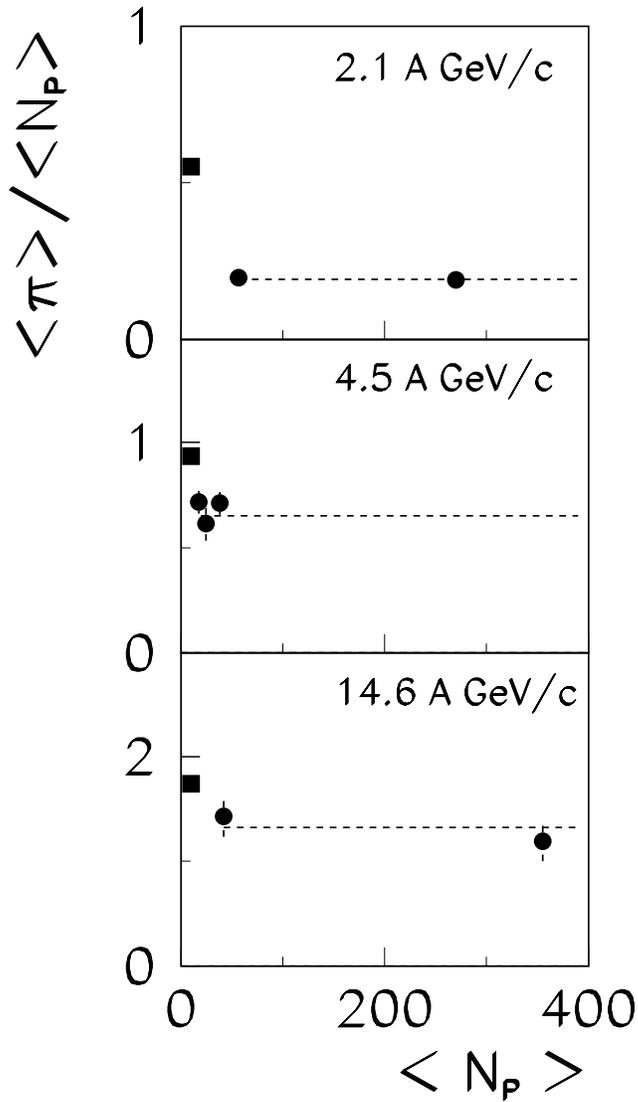}
\caption{
The pion multiplicity per participant nucleon
$\langle \pi \rangle/\langle N_P \rangle$ as a function of the
participant number for nucleon--nucleon interactions (square) and
central collisions of identical nuclei (circles) at 2.1, 4.5, and
15 A$\cdot$GeV/$c$. The lines are shown to guide the eye.
}
\label{fig1}
\end{figure}

\begin{figure}[p]
\epsfig{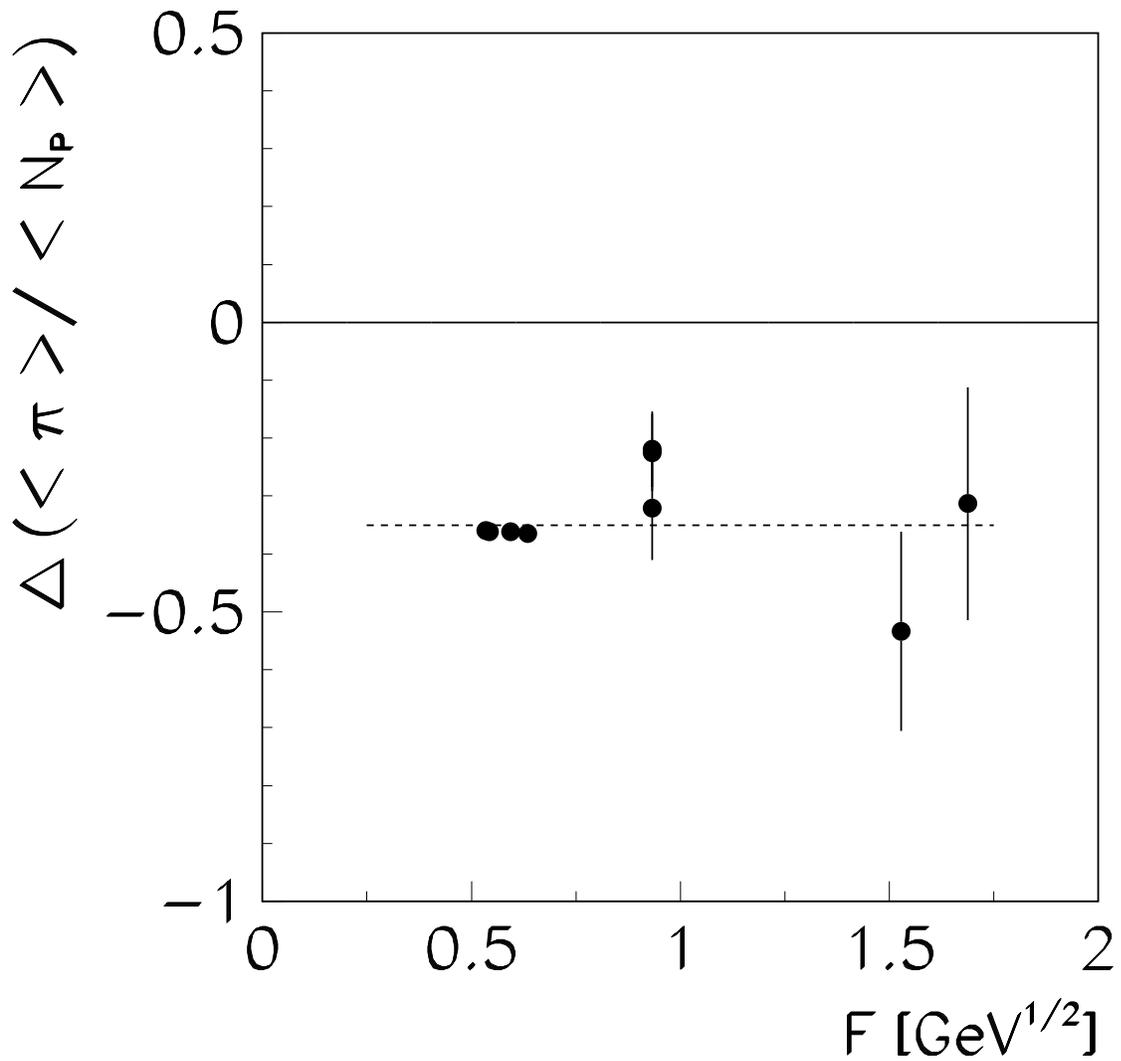}
\caption{
The experimentally measured suppression factor
$\Delta (\langle \pi \rangle / \langle N_P \rangle)$ as function of
the collision energy which is expressed through the Fermi variable
defined as 
$F \protect\equiv (\protect\sqrt{s_{NN}} - 2m_N)^{3/4}/s_{NN}^{1/8}$.
The dashed line shows the mean value equal $-0.35$.
}
\label{fig2}
\end{figure}

\begin{figure}[p]
\epsfig{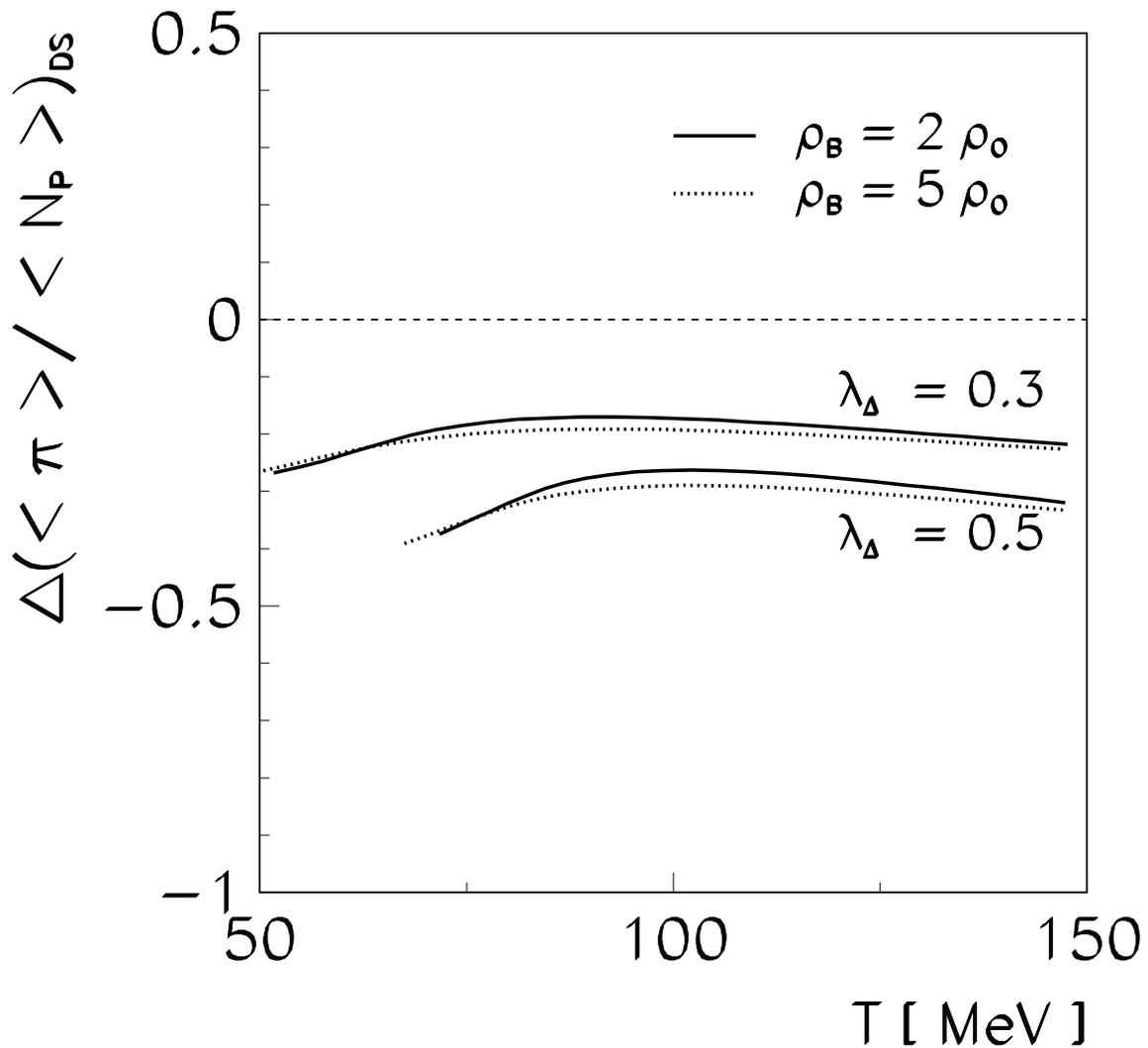}
\caption{
The suppression factor (\protect\ref{supres1}) 
as a function of the temperature $T$.
The extreme cases of $\rho_B = 5 \rho_0$ and $\rho_B = 2 \rho_0$ are
shown by, respectively, the dotted and solid lines. The upper pair of the
dotted and solid  lines corresponds to $\lambda_{\Delta} = 0.3$ while the
lower one to $\lambda_{\Delta} = 0.5$.
}
\label{fig3}
\end{figure}

\begin{figure}[p]
\epsfig{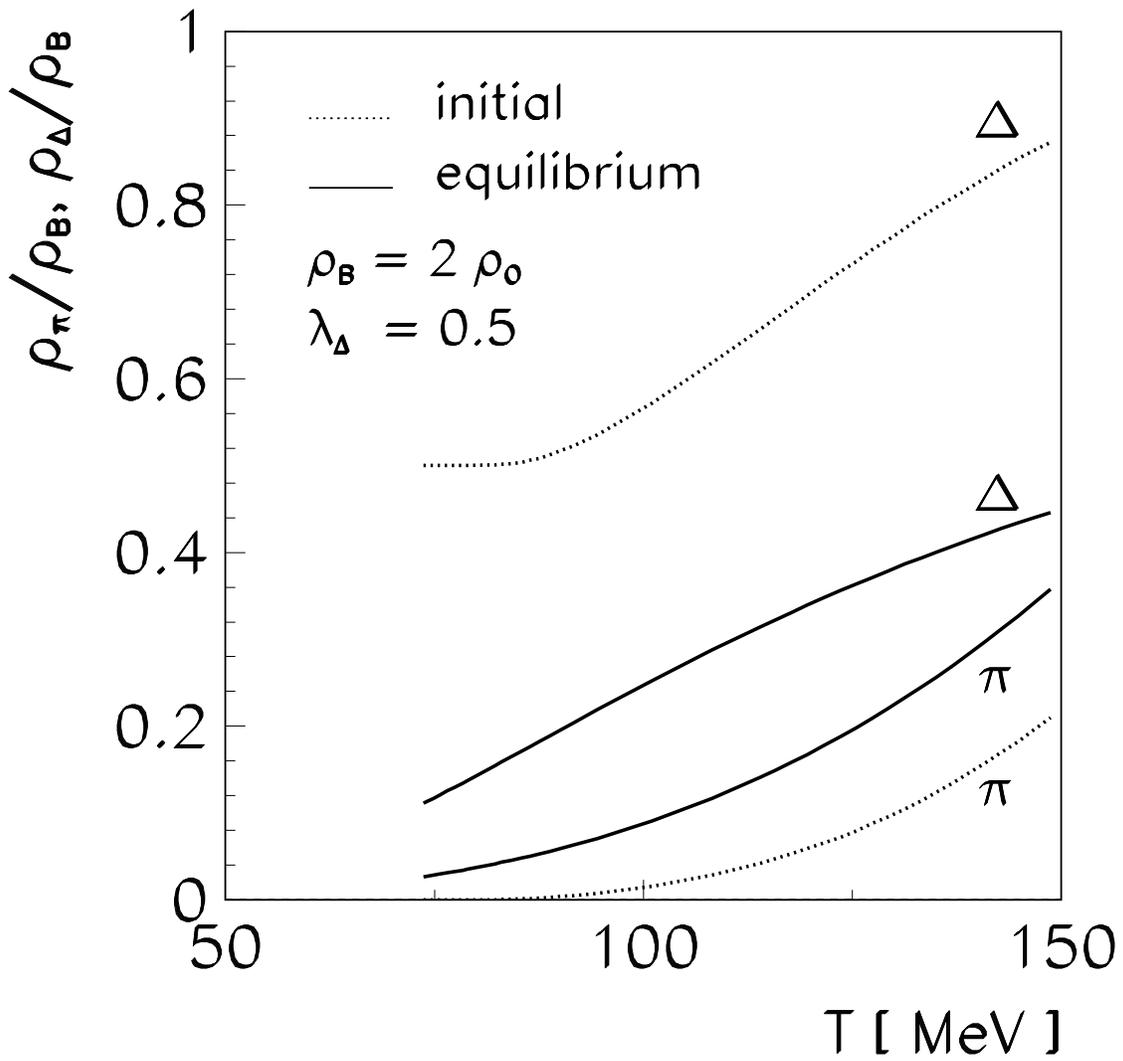}
\caption{
$\rho_{\pi}/\rho_B$ and
$\rho_{\Delta}/\rho_B$ ratios in the initial state (dashed lines) and
in in the chemical equilibrium state (solid lines). The equilibrium baryonic
density is chosen as $2\rho_0$ and $\lambda_{\Delta}=0.5$. 
}
\label{fig4}
\end{figure}

\begin{figure}[p]
\epsfig{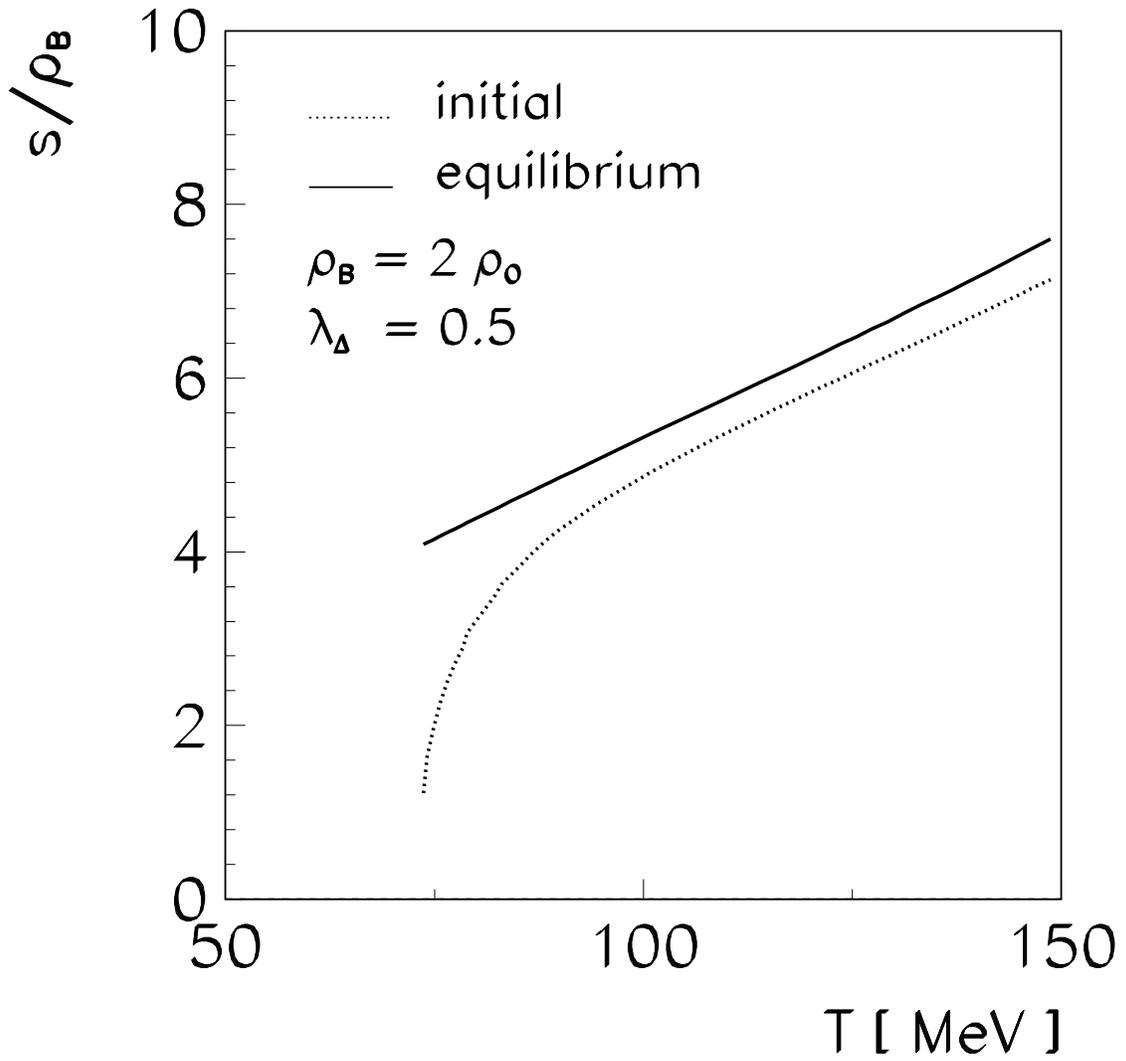}
\caption{
The entropy per participating nucleon
in the initial state (dotted lines) and
in in the chemical equilibrium state (solid lines). The equilibrium baryonic
density is chosen as $2\rho_0$ and $\lambda_{\Delta}=0.5$.
}
\label{fig5}
\end{figure}

\begin{figure}[p]
\epsfig{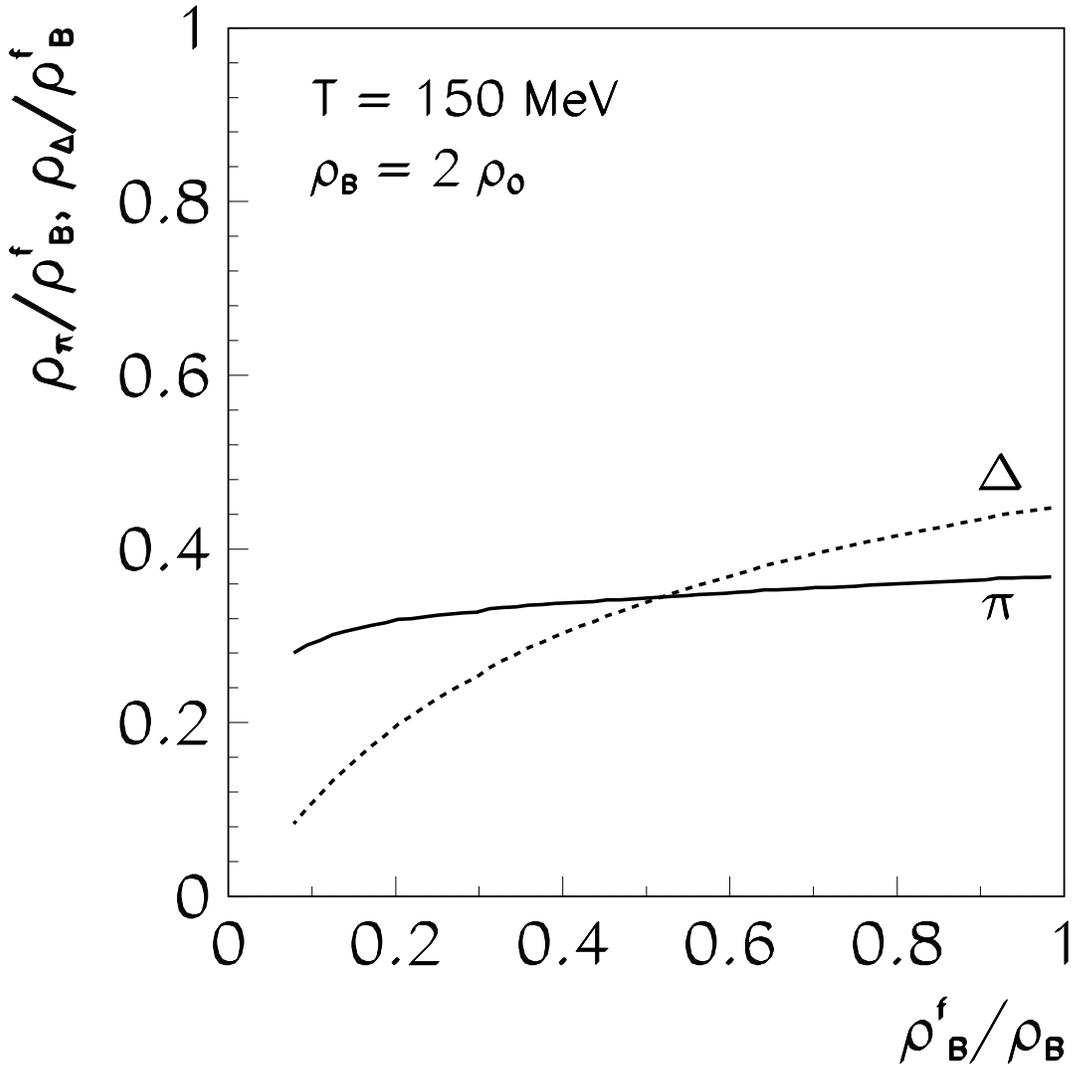}
\caption{
$\rho_{\pi}/\rho_B^f$ (solid line) and 
$\rho_{\Delta}/\rho_B^f$ (dashed line) ratios 
in an isentropic expansion  as a function of $\rho_B^f/\rho_B$.
The parameters are 
$T=150$ MeV and $\rho_B=2\rho_0$.
}
\label{fig6}
\end{figure}

\begin{figure}[p]
\epsfig{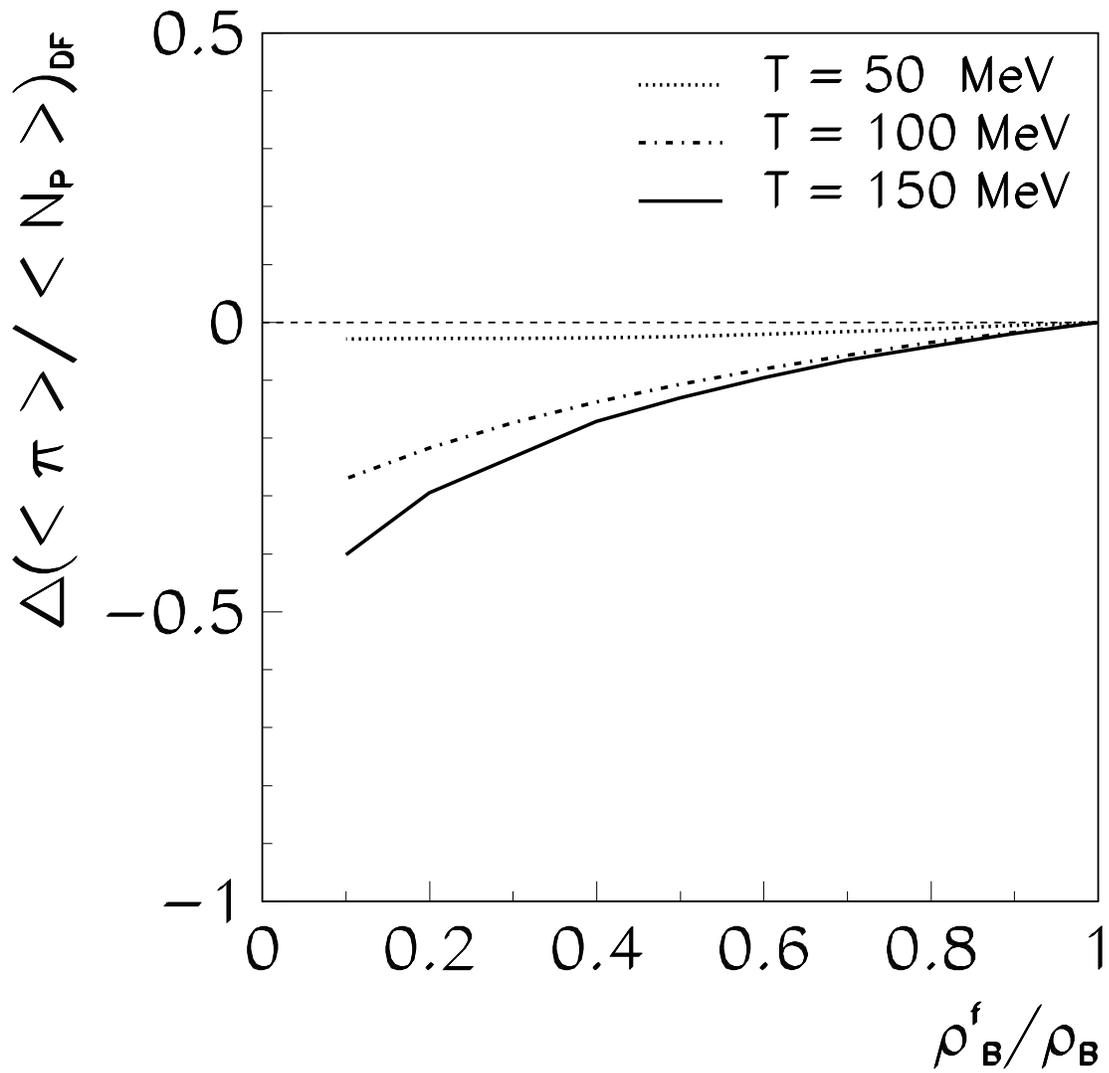}
\caption{
The suppression factor 
(\protect\ref{supres2}) as a function of $\rho_B^f/\rho_B$.
The dotted, dotted--dashed and solid line corresponds to $T$ equal 50, 100
and 150 MeV, respectively.
}
\label{fig7}
\end{figure}

\end{document}